\documentclass[aps,prb,twocolumn,superscriptaddress,reprint,longbibliography]{revtex4-2}
\usepackage{amsmath,amssymb,graphicx,bm}
\usepackage{subcaption}
\usepackage{ulem}
\usepackage[colorlinks=true,linkcolor=blue,citecolor=blue,urlcolor=blue]{hyperref}
\usepackage{tikz}
\usepackage{tikz-feynman}
\usepackage{float}
\usepackage{color}
\usetikzlibrary{decorations.pathmorphing, backgrounds, positioning, arrows.meta,decorations.markings}
\tikzset{
  prop/.style={
    thick,
    decoration={
      markings,
      mark=at position 0.5 with {
        \arrow[xshift=1.5mm]{Latex[length=2.5mm]}
      }
    },
    postaction={decorate}
  }
}

\newcommand{\Rv}{\mathbf{R}}
\newcommand{\kv}{\mathbf{k}}
\newcommand{\qv}{\mathbf{q}}

\begin{document}

\title{Strange metallicity in the Kagome metal Ni$_3$In: a DMFT investigation}

\author{Ruslan Mushkaev}
\affiliation{Department of Physics, University of Fribourg, 1700, Fribourg, Switzerland}

\author{Francesco Petocchi}
\affiliation{Department of Physics, University of Fribourg, 1700, Fribourg, Switzerland}

\author{Philipp Werner}
\affiliation{Department of Physics, University of Fribourg, 1700, Fribourg, Switzerland}

\date{\today}

\begin{abstract}
    Strange metallicity, characterized by a linear temperature dependence of the resistivity, is observed in a broad range of correlated materials, including heavy-fermion compounds and cuprate superconductors. It has also recently been reported for the Kagome metal Ni$_3$In, where almost localized and itinerant electronic degrees of freedom coexist as a result of a partially flat band. We investigate the correlated electronic structure and transport properties of Ni$_3$In with dynamical mean field theory (DMFT) calculations performed on a minimal single-band Hubbard model, constructed from compact molecular orbitals. Despite the large band filling, even for moderate Hubbard repulsion, we observe a non-Fermi-liquid like frequency dependence of the self-energy, as well as the formation of local magnetic moments. With increased hole doping, a crossover to a heavy Fermi-liquid regime is found. We interpret these results in terms of an effective model for the partially filled narrow band near $k_z=0$.     
\vspace{1.5cm}
\end{abstract}

\maketitle

\section{Introduction}

Correlated electron materials with local interactions comparable to the bandwidth display a variety of remarkable phenomena. The interaction effects in these systems are further enhanced if the bands near the Fermi level become partially flat~\cite{sayyad_nfl_one_band}, as it is the case, for example, in hole-doped cuprates near optimal doping. Another prototypical manifestation of such enhanced correlation effects is found in the Kagome metals, where hopping interference on a geometrically frustrated lattice results in a sharply peaked density of states. These (partial) flat band systems host a variety of orders, such as charge density waves~\cite{av3sb5_kagome,fege_cdw,kv3sb5_cdw}, unconventional superconductivity~\cite{thomale_kagome,topology_sc_kagome,sc_cr_based_kagome}, magnetism~\cite{fege_cdw_mag,kagome_magnets_nature} and strange metallicity~\cite{ni3in_nature,ni3in_origin_nature,ni3in_cmo}.

Strange metal states with an almost linear in temperature ($T$) resistivity are observed in several classes of correlated electron systems, often in connection with the emergence of local moments and (at lower $T$) unconventional superconductivity. Prominent examples are heavy-fermion systems~\cite{resistivity_heavy_fermion_nature}, cuprate superconductors~\cite{resistivity_la2cuo4_science}, and magic-angle twisted bilayer graphene~\cite{matbg_strange_metallicity}. A recent addition is the Kagome metal Ni$_3$In~\cite{ni3in_nature,ni3in_origin_nature}, where flat-band-like Ni-$d_{xz}$ states near the Fermi level are believed to play a role similar to that of localized magnetic moments in a typical $f$-electron Kondo lattice system. This scenario is rather unusual, since it means that Kondo-like behavior emerges from the interaction of (almost) localized and delocalized electrons originating from the same band~\cite{ni3in_nature,strange_metals_perspective_nature}, although such a scenario has also been discussed for cuprates~\cite{Werner2016,hidden_kondo_werner}. In contrast, in a conventional heavy-fermion system, a lattice of localized magnetic moments interacts with a conduction electron band, as described by the periodic Anderson model. 

Ni$_3$In exhibits a linear resistivity in the range of $\sim$ 2-100 K, as well as an upturn in the specific heat at low temperatures, which are both signatures of non-Fermi liquid (NFL) like behavior~\cite{ni3in_nature}. Furthermore, the compound is potentially in the vicinity of a quantum critical point (QCP) and may thus be influenced by a transition from an ordered state into the heavily renormalized Fermi-liquid-like (FL) regime at 0 K. The microscopic origins of strange metallicity are still a subject of intense study, with theories ranging from Kondo destruction to SYK-based models~\cite{strange_metals_perspective_nature}.

In this paper, we investigate the correlated electronic structure and transport properties of Ni$_3$In with dynamical mean field theory (DMFT)~\cite{dmft_georges}. We show that the linear-in-$T$ resistivity can be reproduced by a minimal effective single-band model, constructed from compact molecular orbitals (CMOs) which form an effective stacked triangular lattice~\cite{ni3in_cmo}. Non-Fermi liquid properties show up in the local DMFT self-energy, whose imaginary part exhibits a square-root like scaling with the Matsubara frequencies, unlike the expected linear slope for a Fermi-liquid. 
We also study the effect of hole doping, which leads to an increase in the NFL-FL crossover temperature. The analysis of the static, local spin susceptibility indicates the formation of local moments in the undoped compound. Finally, by doubling the unit cell and performing bonding-anti-bonding transformations in the stacking direction, we link the NFL behavior to the bonding electrons, which dominantly contribute to the flat-band weight near $k_z=0$, and speculate about the proximity to a ferromagnetic instability.  

The paper is structured as follows. Section~\ref{sec_methods} discusses the crystal structure of Ni$_3$In, as well as the construction of the compact molecular orbitals, on which single-band DMFT calculations are performed. It also outlines the details behind the resistivity calculations within DMFT. Section~\ref{sec_res} presents the main results in the form of temperature-dependent resistivities and self-energies at various dopings, as well as the local, static spin susceptibility, while Section~\ref{sec_conc} summarizes the main findings. Technical details related to the resistivity calculation are presented in the Appendix.

 \begin{figure*}[t] 
        \centering
        \begin{subfigure}{0.3\textwidth}
            \centering
            \includegraphics[height=0.2\textheight]{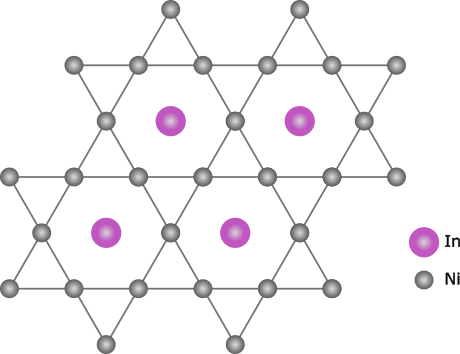}
            \caption{}
            \label{ni3in_layer}
        \end{subfigure}
        \begin{subfigure}{0.3\textwidth}
            \centering
            \includegraphics[height=0.2\textheight]{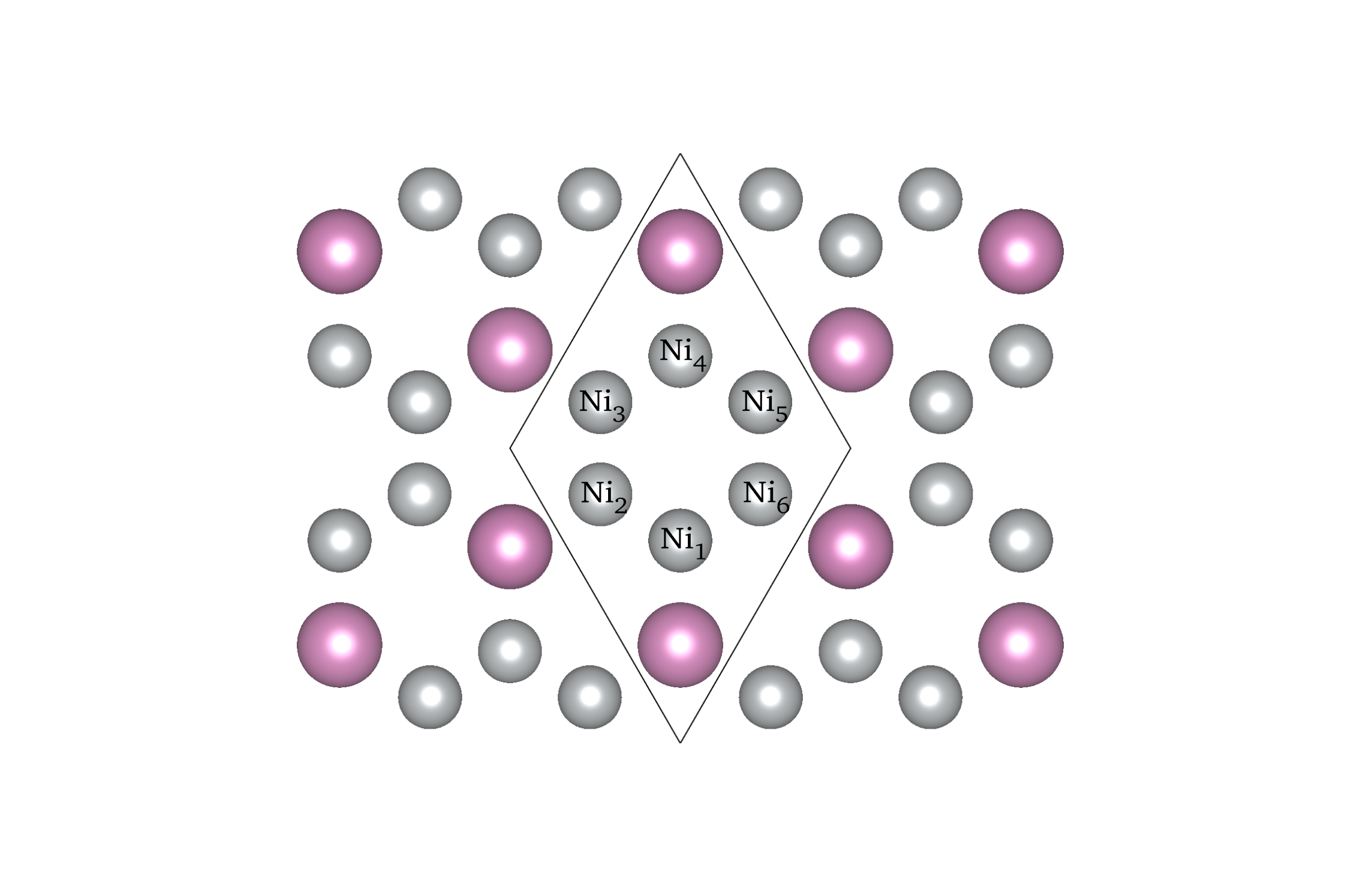}
            \caption{}
            \label{ni3in_star_unit_cell}
        \end{subfigure}
        \begin{subfigure}{0.3\textwidth}
            \centering
            \includegraphics[height=0.2\textheight]{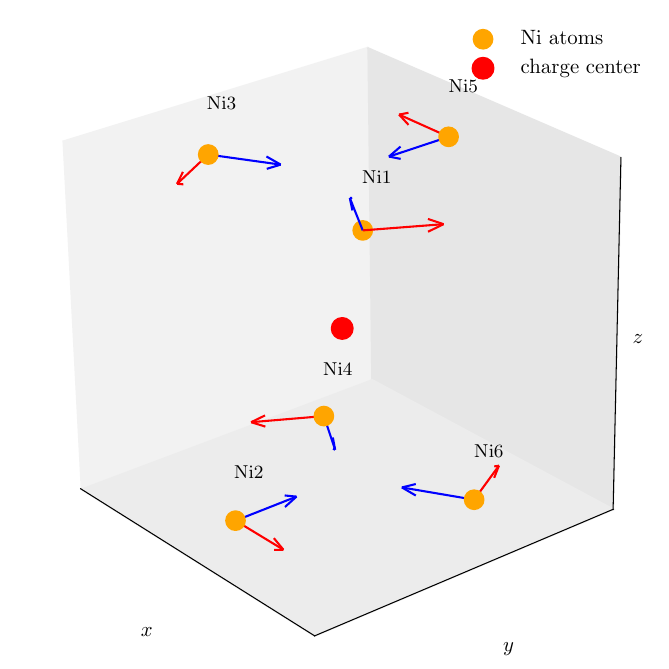}
            \caption{}
            \label{ni3in_local_axes}
        \end{subfigure}
        \caption{Panel (a): one of the Ni$_3$In Kagome layers. Panel (b): shifted Ni$_3$In unit cell. Panel (c): local orbital axes.}
        \label{ni3in_star}
    \end{figure*}

\section{Methods}
\label{sec_methods}

\subsection{Crystal structure and effective single-band model}
\label{sec_crystal_structre_cmo}

The crystal structure of Ni$_3$In is well studied~\cite{ni3in_nature}. Crystallizing in the hexagonal $P6_3/mmc$ space group, it is composed of Ni bilayers, in which each layer forms a breathing Kagome lattice. The In atoms are centered in the middle of the Ni hexagons: see Fig.~\ref{ni3in_star}(a) for a depiction of one of the layers.

The primitive unit cell contains two In atoms (one for each layer), as well as 6 Ni atoms which are AB-stacked (purple and grey, respectively, in Fig.~\ref{ni3in_star}(b)). The atomic positions in units of the lattice vectors are reported in Table~\ref{tab:atomic_positions}. For the Ni atom at the 6h Wyckoff position, we use a value of $x=0.8437$, and the experimental lattice constants $a = 5.335$ \AA, $c = 4.236$ \AA~\cite{ni3in_prx} for the subsequent DFT and Wannier90 calculations.

\begin{table}[b]
\centering
\begin{tabular}{|c|c|c|c|c|}
\hline
\textbf{Element} & \textbf{Wyckoff position} & $\textbf{x}$ & $\textbf{y}$ & $\textbf{z}$ \\
\hline
Ni & 6h & $x$ & $2x$ & 1/4 \\
In & 2c & 1/3 & 2/3 & 1/4 \\
\hline
\end{tabular}
\caption{Ni$_3$In atomic positions in fractional coordinates.}
\label{tab:atomic_positions}
\end{table}

\begin{figure*}[t]
        \centering
        \includegraphics[width=0.9\textwidth]{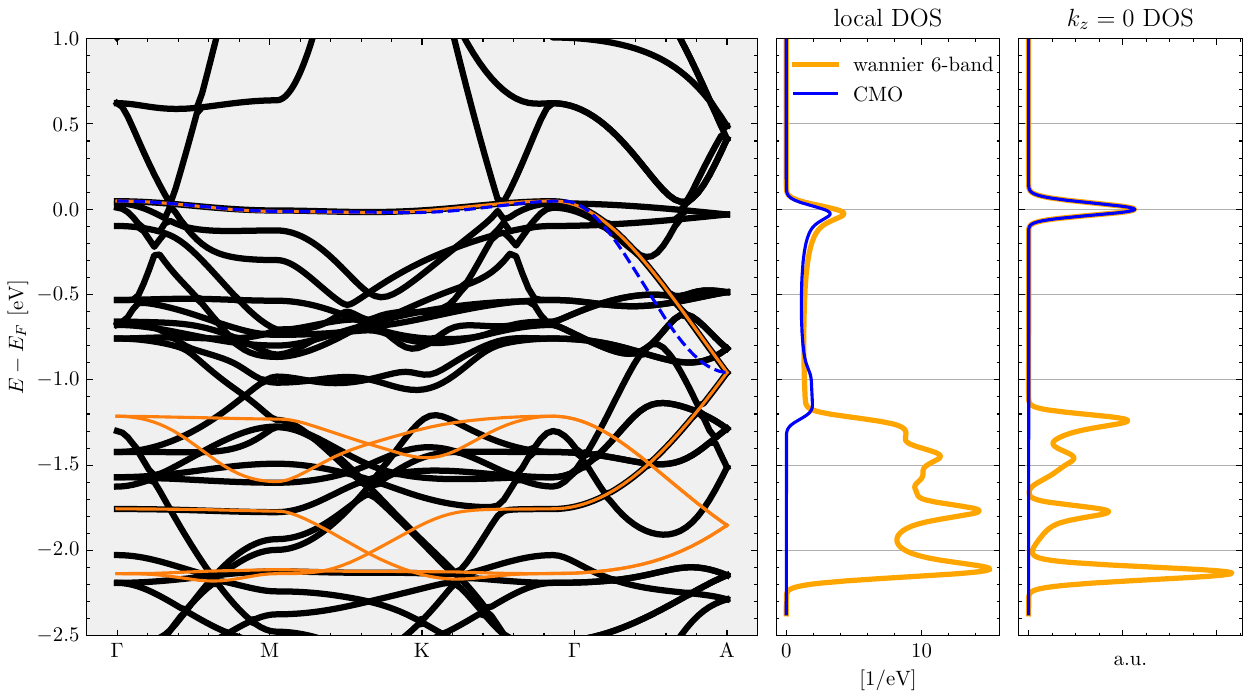}
        \caption{Left panel: DFT bandstructure (black) and 6 Wannier bands (orange), derived from a Ni-$d_{xz}$ Wannier Hamiltonian. The band derived from the CMO states is the blue dashed line.
        Middle and right panels: Full DOS for the 6-band and 1-band (CMO) models, and DOS calculated in the $k_z=0$ plane.
        }
        \label{ni3in_dft_wannier_6band}
\end{figure*}

The DFT calculations reveal that the bands near the Fermi level are predominantly composed of Ni-$d_{xz}$ orbitals in local, rotated coordinates~\cite{ni3in_nature}. Following the approach of Ref.~\cite{ni3in_cmo}, we choose a unit cell composed of two star-stacked Ni triangles, as depicted in Fig.~\ref{ni3in_star}(b). The 6 Ni atoms are related by a six-fold screw symmetry, and we choose the coordinate system such that the local $x$-axes of the Ni$_i$ and Ni$_{i+1}$ atoms differ by a $\pi/3$ rotation (Fig.~\ref{ni3in_star}(c)).

This construction yields 6 Wannier bands, which are depicted by the orange lines in Fig.~\ref{ni3in_dft_wannier_6band}. The band near the Fermi level is well reproduced, while the remaining five bands are much lower in energy, completely filled, and hence not actively participating in the key physics. To isolate the contribution of the top-most band, we project the $6 \times 6$ Wannier Hamiltonian onto the so-called compact molecular orbital state (CMO)~\cite{ni3in_cmo}, which is a local superposition of the 6 Ni-$d_{xz}$ Wannier states in the unit cell:
\begin{equation}
    | \text{CMO} \rangle := \frac{1}{\sqrt{6}} \Big( | d^1_{xz} \rangle -| d^2_{xz} \rangle + | d^3_{xz} \rangle - | d^4_{xz} \rangle + | d^5_{xz} \rangle - | d^6_{xz} \rangle \Big),
\end{equation}
where the superscript indexes the Ni atoms. Such a state was shown in Ref.~\cite{ni3in_cmo} to reproduce the flat band, using an elementary band representation analysis. Indeed, projecting our Hamiltonian onto the CMO state results in a single band which is in good agreement with the almost flat band in the $k_z=0$ plane (blue dashed line in Fig.~\ref{ni3in_dft_wannier_6band}). The deviation along the $\Gamma$-$A$ line is likely due to the CMO state having some weight on the lower branch of the neighboring orange band, as can be seen from the supplemental material in Ref.~\cite{ni3in_cmo}. The CMO states are localized at the centers of the star-shaped plaquettes (see Fig.~\ref{ni3in_star}(c)), which form a stacked triangular lattice. Integration of the density of states yields a filling of $n \simeq 1.85$ electrons for the CMO band. 

In the following, we set $\hbar=e=V_\text{u.c.}=1$, where $V_\text{u.c.}$ is the unit cell volume.

\subsection{Many-body calculation}

Using this effective single-orbital model on a stacked triangular lattice, we perform paramagnetic single-site DMFT calculations \cite{dmft_georges}, using a hybridization-expansion continuous-time Monte Carlo impurity solver \cite{Werner2006}.  We choose local Hubbard interactions of $U=1,2$ eV, which are comparable to the bandwidth of the single band, and inverse temperatures $\beta$ in the range of 60-360 eV$^{-1}$, corresponding to temperatures of 32-193 K. The DMFT self-consistency loop employs a $16\times16\times64$ $k$-point grid.

\subsection{Resistivity calculation}
\label{sec_res_dmft}

The temperature-dependent resistivity, $\rho$, is computed within the DMFT approximation. We start from the current-current correlation function, $\Lambda$, which is a simple bubble within DMFT, due to vanishing vertex corrections~\cite{dmft_georges} (the vertex in DMFT is local and the band velocities satisfy $v_{-\kv} = -v_\kv$).  Expressed in momentum space it reads:
\begin{align}
    & \Lambda_\text{plane}(\qv,i\nu_m) = \nonumber\\
    & - \frac{1}{N_\kv \beta} \sum_{\substack{\kv,\alpha \\ i\omega_n,\sigma}} (v_\kv^{\alpha})^2 G_\sigma(\kv,i\omega_n) G_\sigma(\kv+\qv,i\omega_n+i\nu_m) ,
\end{align}
where $i\omega_n$ and $i\nu_m$ are fermionic and bosonic Matsubara frequencies, respectively, and $v_\kv = \partial_\kv \varepsilon_\kv$ is the band velocity. Furthermore, since we are interested in the in-plane conductivity, in order to compare our numerics with experimental data in Ref.~\cite{ni3in_nature}, we compute an average over the $\alpha = x,y$ components. \newline

Given this in-plane correlation function, one obtains the static, uniform, in-plane conductivity $\sigma_\text{dc}$ via~\cite{conductivity_hubbard_prl}
\begin{equation}
    \sigma_\text{dc,plane} = -\partial_{\nu_m} \text{Re} \Lambda_\text{plane}(\qv=0,i\nu_m)|_{\nu_m \rightarrow 0^+} .
    \label{sigma_dc}
\end{equation}
The static conductivity is related to the dc resistivity by $\rho_\text{dc,plane} = 1/\sigma_\text{dc,plane}$. 

In practice, it is difficult to evaluate the dc resistivity from a finite-difference derivative, especially at high temperatures, since the spacing between the neighboring Matsubara points becomes large. To compute the derivative more reliably, one can interpolate the current-current correlator for the first few frequencies. This requires knowledge of $\Lambda(i\nu_m)$ in the vicinity of $i\nu_m = 0$. In the metallic regime, the low-frequency optical conductivity on the real frequency axis can be modeled by a Drude-like form, 
\begin{equation}
    \sigma(\omega) = \frac{D}{\Gamma - i \omega},
\end{equation}
where $D$ is the Drude weight and $\Gamma$ the scattering rate. The conductivity itself is related to the current-current correlator by:
\begin{equation}
    \sigma(\omega) = \frac{\Lambda(\omega)-\Lambda(\omega=0)}{i \omega} .
\end{equation}
In Matsubara space, we thus have the relation 
\begin{equation}
    \Lambda(i\nu_m) = \Lambda(0) - \nu_m \sigma(i\nu_m) = \Lambda(0) - \frac{\nu_m D}{\Gamma + \nu_m} .
\end{equation}

\begin{figure}[H] 
    \centering
        \includegraphics[width=\linewidth]{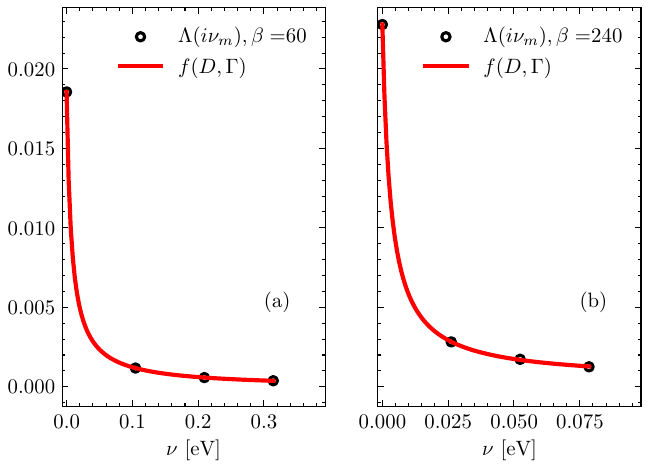}
        \caption{Drude fit of the current-current correlation function for (a) $\beta=$60 and (b) 240 eV$^{-1}$ (193 K and 48 K respectively).
        }
        \label{lambda_fit}
\end{figure}

Using Eq.~\eqref{sigma_dc}, the expression for the dc conductivity becomes $\sigma_\text{dc} = \frac{D}{\Gamma}$. For a given temperature, we hence fit the current-current correlator with the fitting function $f(D,\Gamma) = \Lambda(0) - \frac{\nu_m D}{\Gamma + \nu_m}$ to the first few discrete Matsubara frequencies to obtain a set of optimal parameters $(D,\Gamma)$, which are then used to compute the conductivity and resistivity. Two examples of this Drude fit to the first four frequencies are shown in Fig.~\ref{lambda_fit}.

The above approach has been tested by comparing the numerical results to the analytical form of the $T^2$-resistivity for the quarter-filled square-lattice Hubbard model in the weak coupling regime, see Appendix~\ref{res_benchmark} for more details.

\section{Results}
\label{sec_res}

\subsection{Spectral functions and self-energies}

\begin{figure}[b]
        \centering
        \includegraphics[width=0.4\textwidth]{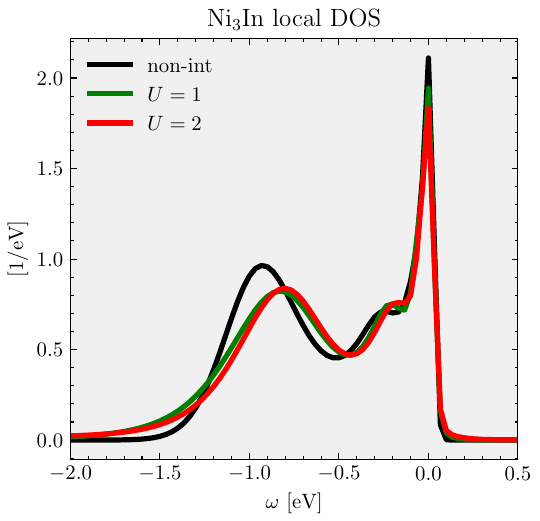}
        \caption{DMFT local density of states per spin for the one-band model of Ni$_3$In for $U=1$ eV and 2 eV, at $\beta=100$ eV$^{-1}$ (116 K). The black curve is the local DOS of the noninteracting model.}
        \label{ni3in_local_dos}
\end{figure}

The local density of states (DOS) obtained from DMFT for the one-band model of Ni$_3$In at $\beta=$ 100 eV$^{-1}$ (100~K) is depicted in Fig.~\ref{ni3in_local_dos}. In the interaction range considered, we find only a very slight narrowing of the DOS and quasiparticle band with increasing $U$. Since the total band filling of $n=1.85$ is large, one might naively expect weak electronic correlations. 

\begin{widetext}
    
\begin{figure}[H]
    \centering
    \begin{subfigure}{0.9\textwidth}
        \centering
        \includegraphics[width=\linewidth]{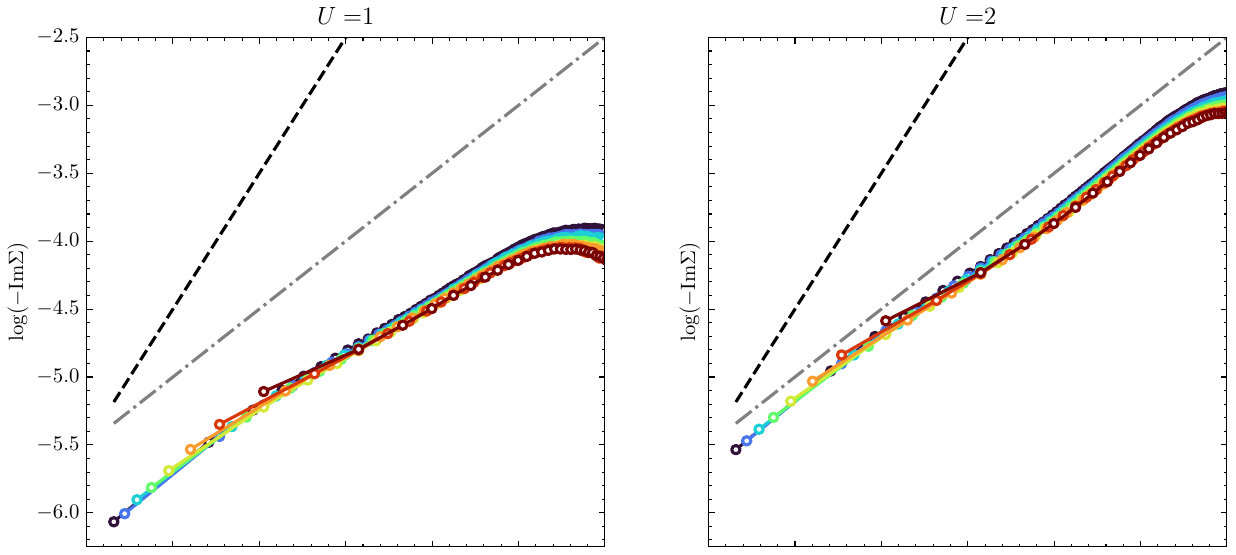}
        \vspace{0.3cm}
    \end{subfigure}
    \begin{subfigure}{0.9\textwidth}
        \centering
        \includegraphics[width=\linewidth]{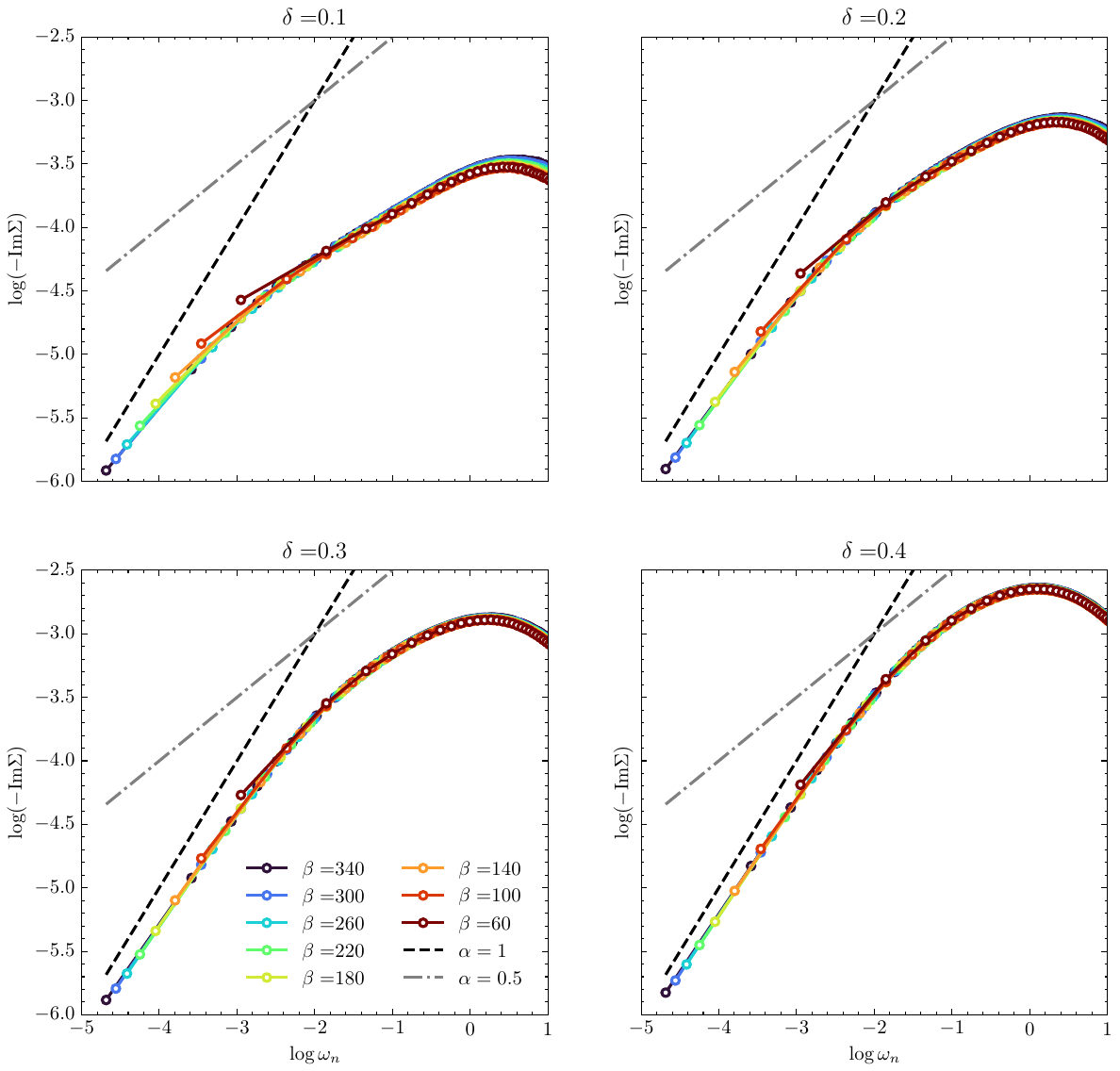}
    \end{subfigure}
    \caption{Local DMFT self-energies for Ni$_3$In for different interactions and temperatures in the undoped case (top row) and for different doping values $\delta = 1.85 - n$ at $U=1$ eV (second and third rows).
    }
    \label{simp_delta_u_ni3in}
\end{figure}

\end{widetext}

However, as shown in the top row of Fig.~\ref{simp_delta_u_ni3in}, the imaginary part of the self-energy exhibits a square-root like dependence on the Matsubara frequency (dot-dashed line), in stark contrast to the linear behavior in a Fermi liquid state (dashed line). We note that this square-root-like scaling is particularly clear at high temperatures and high frequencies. As the temperature is lowered, the slope at the lowest Matsubara frequencies is expected to tend towards 1, marking the crossover to a heavily renormalized Fermi liquid. However, according to the experiments of Ref.~\cite{ni3in_nature}, this crossover should occur at a very low temperature, $T \sim 2$-$3$ K, which is beyond the reach of our DMFT calculations.  

The non-Fermi liquid behavior is apparently related to the presence of the almost flat, partially filled band in the $k_z=0$ plane. Similar observations of NFL behavior associated with sharp DOS features have been reported in simulations of pyrochlore compounds~\cite{prl_nfl_shinaoka} and in model studies of partial flat-band systems~\cite{Sayyad2020}.  

We next investigate the effect of doping on the self-energy, which is illustrated for $U=1$ in the second and third rows of Fig.~\ref{simp_delta_u_ni3in}. Hole-doping the single band away from the DOS peak is expected to alter the NFL-FL crossover temperature. We define the doping parameter $\delta = 1.85 - n$, where 1.85 is the nominal electron filling in the undoped case and $n$ the filling of the model.
Hole doping changes the slope of the self-energy at the lowest accessible temperature. It approaches 1 for dopings $\delta \geq 0.2$, marking the crossover to the Fermi-liquid regime. Hence, the NFL-FL crossover temperature is significantly enhanced with increasing hole-doping.

\subsection{DMFT resistivity}

\begin{figure}[t]
        \centering
        \includegraphics[width=0.42\textwidth]{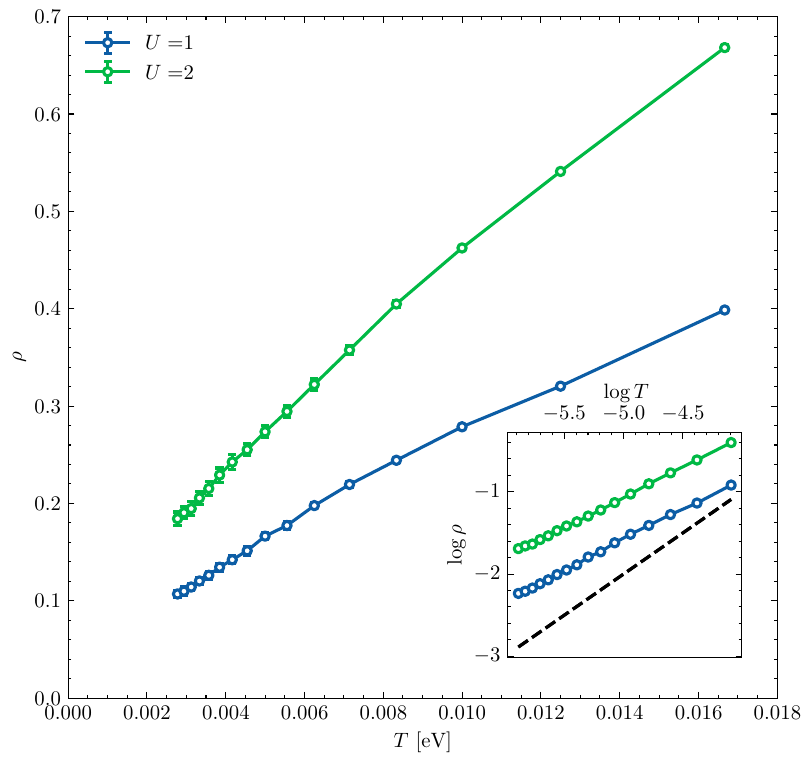}
        \caption{In-plane resistivity $\rho(T)$ for Ni$_3$In for $U=1,2$ eV on a linear scale and on a log-log scale (inset). The black dashed line is a linear curve.
        }
        \label{rho_t_undoped}
\end{figure}

The temperature dependent in-plane resistivity $\rho$ is depicted in Fig.~\ref{rho_t_undoped}, for the undoped model with interactions $U=1$ and $2$ eV. The (very small) vertical error bars correspond to the numerical uncertainty of the fitting procedure, see Appendix~\ref{uncertainty_fit} for a detailed discussion. The resistivity behaves sub-linearly in a wide temperature range, $\rho\propto T^\alpha$, with an exponent $\alpha \simeq 0.75$, which is a signature of the NFL regime. For comparison, the black dashed line in the log-log plot shows the linear scaling ($\alpha=1$). Increasing the value of $U$ does not qualitatively change the observed trend, but results in an overall enhancement of the resistivity for all temperatures, due to heavier charge carriers. As mentioned above, for the nominal filling $n=1.85$, the crossover to the heavily renormalized FL is expected to occur for much lower temperatures, of the order of $\sim 1$~K, which is beyond the reach of our calculations. 

\begin{figure}[H]
    \centering
    \includegraphics[width=0.4\textwidth]{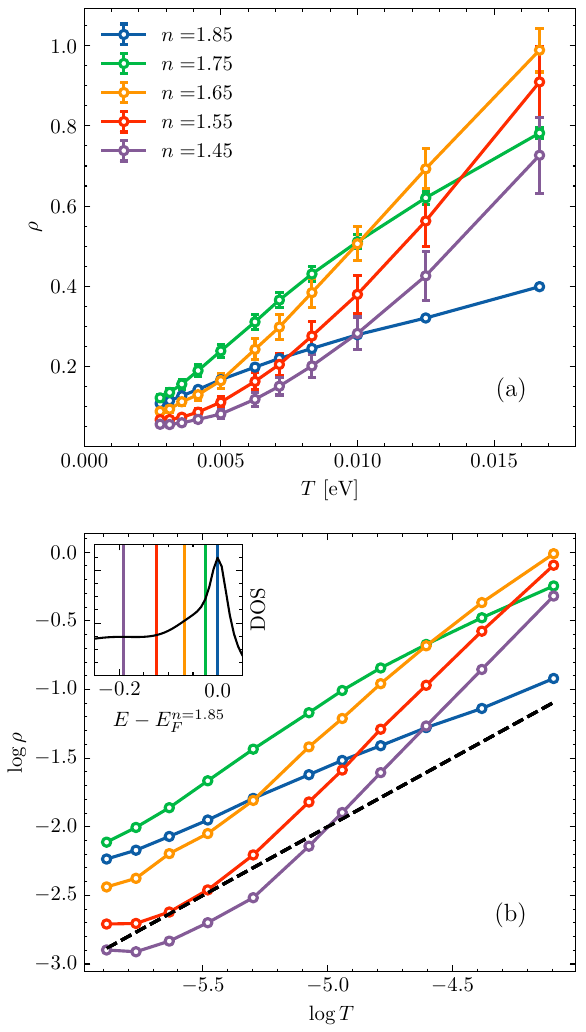}
    \caption{In-plane resistivity $\rho(T)$ for $U=1$ eV, and different hole dopings. Panels (a) and (b) show the results on linear and log-log scales respectively. Inset: position of the chemical potential relative to the DOS at $\beta=100$ eV$^{-1}$ (116 K). 
    }
    \label{rho_t_doped}
\end{figure}

The resistivity is plotted for different values of hole dopings away from the initial $n \simeq 1.85$ electron filling in Fig.~\ref{rho_t_doped}. The inset in the log-log panel depicts the position of the chemical potential in relation to the $U=1$ eV DOS at $\beta=100$ eV$^{-1}$ (116 K). We note that for fillings $n=1.85$ and $n=1.75$, the chemical potential is in the vicinity of the DOS peak. For these fillings, the resistivity exhibits a clear NFL scaling in a broad temperature window, with $\rho \sim T^{0.75}$ and $\rho \sim T^1$, respectively.

\begin{figure}[H]
    \centering
    \includegraphics[width=0.5\textwidth]{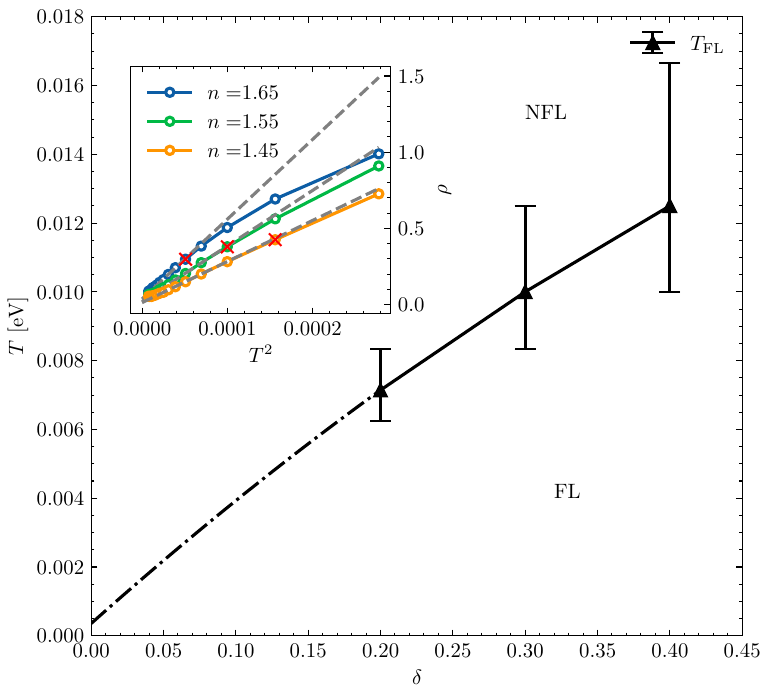}
    \caption{Temperature-doping phase diagram with the high-temperature NFL-FL crossover marked by a black line. The black dashed line is a parabolic extrapolation. Inset: $T^2$ low-temperature fit of the resistivity for different hole dopings $\delta = 1.85-n$. Red crosses denote the onset of FL behavior. 
    }
    \label{rho_t2_fit_t_fl}
\end{figure}

As seen in the plots of the local self-energy (Fig.~\ref{simp_delta_u_ni3in}), increased doping results in a higher NFL-FL crossover temperature. Consistent with this, the quadratic scaling of the resistivity at low $T$ becomes visible for fillings of $n=1.65$ and below. Since the quadratic coefficient of the resistivity in the FL regime scales with the density of states at the Fermi level, $\text{DOS}(E_F)$, the resistivity parabolas at low temperatures become flatter as the hole doping is increased. The resistivity appears to saturate to a small nonzero value as $T \rightarrow 0$, which could be due to the crude nature of the two-parameter fit, or due to the true Drude peak becoming narrower faster than the Matsubara spacing at low temperatures.
We can nevertheless extract some information from the temperature region where the resistivity scales approximately as $\sim T^2$. For this, we plot $\rho$ vs $T^2$ in the inset panel of Fig.~\ref{rho_t2_fit_t_fl}, and perform low-temperature linear fits (discarding the first $\simeq 5$ frequencies) to obtain an estimate of the crossover temperature, $T_\text{FL}$, depicted by the solid line in the quantum-critical like phase diagram in the main panel of Fig.~\ref{rho_t2_fit_t_fl}. $T_\text{FL}$ grows with hole doping $\delta$, becoming slightly flatter as $\delta$ is increased. The dot-dashed line is a qualitative estimate of the phase boundary (obtained from a parabolic extrapolation), since our temperature range is insufficient to reliably resolve $T_\text{FL}$ for $\delta \leq 0.1$. The overall trend agrees with the hypothetical phase diagram proposed in~Ref.~\cite{ni3in_nature}.

\subsection{Static magnetic susceptibility}

DMFT also allows us to measure the local spin correlation function, defined on the Matsubara axis as
\begin{equation}
     \chi_{zz}(\tau) = \langle S_z(\tau) S_z(0) \rangle,
\end{equation}
with $S_z = \frac{1}{2}(n_\uparrow - n_\downarrow)$. The static value of the spin-spin correlation function, $\chi_{zz}^\text{static} = \int_{0}^{\beta} d\tau \chi_{zz}(\tau)$, probes the presence of local moments, which typically yield a Curie-Weiss scaling at high temperatures. 

\begin{figure}[ht!]
        \centering
        \includegraphics[width=0.4\textwidth]{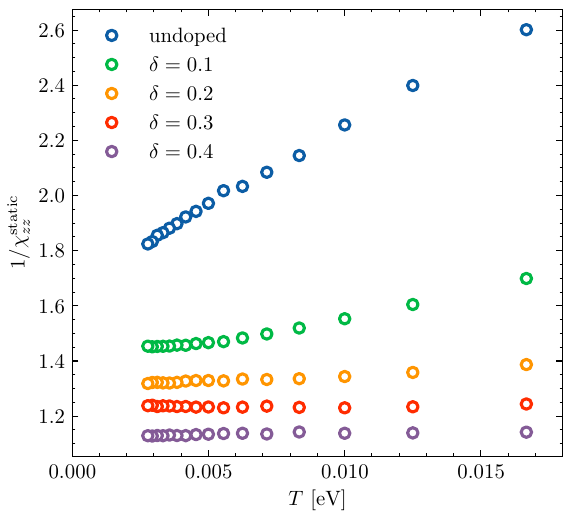}
        \caption{Temperature dependence of the inverse magnetic susceptibility for different doping values at $U=1$ eV.
        }
        \label{chi_loc_m}
\end{figure}

The temperature dependence of the static, local magnetic susceptibility is depicted in Fig.~\ref{chi_loc_m} at various dopings. We note two key trends. Firstly, the presence of local moments gets enhanced as we increase hole doping, manifested by larger values of $\chi_{zz}$. This is because the system is getting closer to half-filling, where the expectation value of $S_z^2$ is largest. Secondly, we observe a Curie-Weiss-like growth of $\chi_{zz}$ at high temperatures, which saturates to a constant Pauli-like value at lower temperatures for dopings $\delta \geq 0.1$. A temperature-independent value of the local spin susceptibility is a signature of the FL regime~\cite{local_spin_dyn_mit,sampling_ct_hyb_kowalski}. In the undoped case, such a saturation is not yet observed, since we are still far above the corresponding NFL-FL crossover temperature. The observed NFL behavior in Ni$_3$In in a wide temperature range is hence linked to the presence of local magnetic moments, which remain unscreened down to a very low temperature. 

\subsection{Stacked triangular lattice model}

Strange metallicity generically occurs in multi-orbital systems with a large Hund's coupling $J$~\cite{nfl_nature_werner}, due to the freezing of local moments in the metallic state, as the system approaches the high-spin Mott phase~\cite{hidden_kondo_werner}. In the present study, however, we are dealing with a single-orbital model, which exhibits similar NFL characteristics. Such a single-orbital system can be mapped onto an effective two-band Hubbard model with Hund's coupling $J=U/2$, by performing bonding-antibonding ($c/f$) transformations within an enlarged unit cell~\cite{hidden_kondo_werner}. If one of the orbitals is close to a half-filled Mott-insulating state, while the other is more weakly correlated and itinerant, the system can furthermore be mapped onto a ferromagnetic Kondo lattice, where strange metallicity is prevalent. 

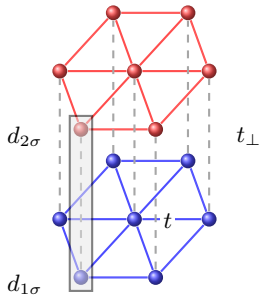
\begin{figure}[b]
    \centering
\resizebox{0.15\textheight}{!}{\begin{tikzpicture}[
    x={(1cm,0cm)},
    y={(0.25cm,0.9cm)}, 
    z={(0cm,0.9cm)},
    site/.style={
        circle,
        inner sep=2pt,
        shading=ball
    },
    hop/.style={thick},
    vert/.style={thick,dashed,gray!70},
    label/.style={fill=white,inner sep=1pt}
]

\def\h{2.2}
\def\a{1}
\def\b{0.866} 

\node[site,ball color=blue!70] (b0) at (0.5*\a,\b,0) {};

\node[site,ball color=blue!70] (b1) at (0,0,0) {};
\node[site,ball color=blue!70] (b2) at (\a,0,0) {};
\node[site,ball color=blue!70] (b3) at (1.5*\a,\b,0) {};
\node[site,ball color=blue!70] (b4) at (\a,2*\b,0) {};
\node[site,ball color=blue!70] (b5) at (0,2*\b,0) {};
\node[site,ball color=blue!70] (b6) at (-0.5*\a,\b,0) {};

\draw[hop,blue!70]
  (b1)--(b2)--(b3)--(b4)--(b5)--(b6)--(b1);

\foreach \i in {1,...,6} {
    \draw[hop,blue!70] (b0)--(b\i);
}

\node[label] at (0.95,0.85,0) {$t$};

\node[site,ball color=red!70] (t0) at (0.5*\a,\b,\h) {};

\node[site,ball color=red!70] (t1) at (0,0,\h) {};
\node[site,ball color=red!70] (t2) at (\a,0,\h) {};
\node[site,ball color=red!70] (t3) at (1.5*\a,\b,\h) {};
\node[site,ball color=red!70] (t4) at (\a,2*\b,\h) {};
\node[site,ball color=red!70] (t5) at (0,2*\b,\h) {};
\node[site,ball color=red!70] (t6) at (-0.5*\a,\b,\h) {};

\draw[hop,red!70]
  (t1)--(t2)--(t3)--(t4)--(t5)--(t6)--(t1);

\foreach \i in {1,...,6} {
    \draw[hop,red!70] (t0)--(t\i);
}

\foreach \i in {0,...,6} {
    \draw[vert] (b\i)--(t\i);
}

\node[label] at (2.,1.0,1.1) {$t_\perp$};

\node[label] at (-1,1.0,1.1) {$d_{2\sigma}$};
\node[label] at (-1,1.0,-1.1) {$d_{1\sigma}$};

\fill[blue!10,opacity=0.35]
  (b1)--(b2)--(b3)--(b4)--(b5)--(b6)--cycle;

\fill[red!10,opacity=0.35]
  (t1)--(t2)--(t3)--(t4)--(t5)--(t6)--cycle;

\def\dx{0.15} 

\coordinate (r1) at ($(b1)+(-\dx,0,-0.2)$);
\coordinate (r2) at ($(b1)+(\dx,0,-0.2)$);
\coordinate (r3) at ($(t1)+(\dx,0,0.2)$);
\coordinate (r4) at ($(t1)+(-\dx,0,0.2)$);

\filldraw[
    fill=gray!20,
    opacity=0.5,
    draw=black,
    thick
] (r1)--(r2)--(r3)--(r4)--cycle;

\end{tikzpicture}
}
\caption{Stacked triangular lattice model for Ni$_3$In. The gray box indicates the two-site unit cell of the transformed model with $c$, $f$ orbitals.}
\label{stacked_triangle_uc}
\end{figure}

Our CMO model is defined on a stacked triangular lattice, characterized by the lattice vectors $\mathbf{a_1} = (a,0,0),\mathbf{a_2} = a(-\frac{1}{2},\frac{\sqrt{3}}{2},0)$ and $\mathbf{a_3} = (0,0,c)$. The effective nearest-neighbor tight-binding Hamiltonian is parameterized by the in-plane and vertical hoppings $t > 0$ and $t_\perp > 0$ respectively, as shown in Fig.~\ref{stacked_triangle_uc}, yielding the dispersion
\begin{equation}
    \epsilon(\kv) = \epsilon_\text{plane}(\kv) + 2 t_\perp \cos(\kv \cdot \mathbf{a_3}),
\end{equation}
where the in-plane contribution reads
\begin{equation}
    \epsilon_\text{plane}(\kv) = 2t \Big[ \cos(\kv \cdot \mathbf{a_1}) + \cos(\kv \cdot \mathbf{a_2}) + \cos(\kv \cdot (\mathbf{a_1}+\mathbf{a_2}))\Big].
\end{equation}

In the case of Ni$_3$In, we have $t_\perp \gg t$, as evidenced by the quasi-flat band in the $k_z=0$ plane, and a widely dispersing portion along the $k_z$ direction, effectively making the system 1D, with strong hoppings along vertically aligned chains and a small horizontal coupling, which can be considered as a perturbation. We choose a vertical unit cell, composed of two sites, depicted as a shaded rectangle in Fig.~\ref{stacked_triangle_uc}. The two-site Hamiltonian reads
\begin{align}
    &H = \nonumber\\
    &\sum_{\kv \in \text{BZ'}, \sigma} \begin{pmatrix} d_{1\kv\sigma}^\dag \\ d_{2\kv\sigma}^\dag\end{pmatrix}^T \!\! \begin{pmatrix}
    \epsilon_\text{plane}(\kv) & t_\perp (1+e^{2i\kv \mathbf{a_3}}) \\ t_\perp (1+e^{-2i\kv \mathbf{a_3}}) & \epsilon_\text{plane}(\kv) \end{pmatrix} \!\! \begin{pmatrix}
        d_{1\kv\sigma} \\ d_{2\kv\sigma}
    \end{pmatrix},
\end{align}

where the momentum is now restricted to a halved Brillouin zone, denoted with a primed index.

Defining the bonding-antibonding transformation by
\begin{align}
    c_{\kv\sigma} = \frac{1}{\sqrt{2}}(d_{1 \kv\sigma} + d_{2 \kv \sigma}) , \quad
    f_{\kv\sigma} = \frac{1}{\sqrt{2}}(d_{1 \kv \sigma} - d_{2 \kv \sigma}), 
    \label{ba_transform}
\end{align}
the transformed Hamiltonian in the $c,f$ basis reads
\begin{equation}
 H = \sum_{\kv \in \text{BZ'} \sigma} (c_{\kv\sigma}^\dag,f_{\kv\sigma}^\dag) \mathcal{H}(\kv)\begin{pmatrix}
        c_{\kv\sigma} \\ f_{\kv\sigma}
    \end{pmatrix},
\end{equation}
where
\begin{equation}
    \begin{aligned}
        \mathcal{H}(\kv) =& \, \varepsilon_\text{plane}(\kv)\mathbb{I} \\ &+ 
        \begin{pmatrix}
        t_\perp\big[1+\cos(2\kv \mathbf{a_3})\big]
        &
        - i t_\perp \sin(2\kv \mathbf{a_3})
        \\[6pt]
        i t_\perp \sin(2\kv \mathbf{a_3})
        &
        - t_\perp \big[1+\cos(2\kv \mathbf{a_3})\big]
        \end{pmatrix} .
    \end{aligned}
\end{equation}

We now compute the $c/f$ orbital character of the original band, as well as the corresponding local DOS contributions. For this, we write down the Fourier transform of the single-site Hamiltonian, and decompose it into a sum over the sublattices:
\begin{equation}
    \begin{aligned}
        d_{\kv} &= \frac{1}{\sqrt{N_I}} \sum_{i} \frac{1}{\sqrt{2}} \sum_{\mathbf{\delta}} d_{\Rv_i+\mathbf{\delta}} e^{i\kv(\Rv_i+\mathbf{\delta})}
        \\ &= \frac{1}{\sqrt{ 2 N_I}} \sum_{i} e^{i\kv\Rv_i} (d_{\Rv_i 1}+d_{\Rv_{i} 2} e^{i\kv\mathbf{a_3}}).
    \end{aligned}
\end{equation}
Applying the bonding-antibonding transformation defined in Eq.~\eqref{ba_transform} yields
\begin{equation}
    d_\kv = \frac{1}{\sqrt{N_I}} \frac{1}{2} \sum_i e^{i\kv\Rv_i} \Big[ c_{\Rv_i}(1 + e^{i\kv\mathbf{a_3}}) + f_{\Rv_i}(1 - e^{i\kv\mathbf{a_3}}) \Big].
\end{equation}
\newline
The corresponding $c,f$ weights are the squares of the amplitudes of the Fourier coefficients:
\begin{equation}
    \begin{aligned}
        w_c(\kv) = \frac{1}{4} |1 + e^{i\kv\mathbf{a_3}}|^2 = \frac{1}{2}(1 + \cos(\kv\mathbf{a_3})), \\ 
        w_f(\kv) = \frac{1}{4} |1 - e^{i\kv\mathbf{a_3}}|^2 = \frac{1}{2} (1 - \cos(\kv\mathbf{a_3})).
    \end{aligned}
\end{equation}

\begin{figure}[t]
        \centering
        \includegraphics[width=0.5\textwidth]{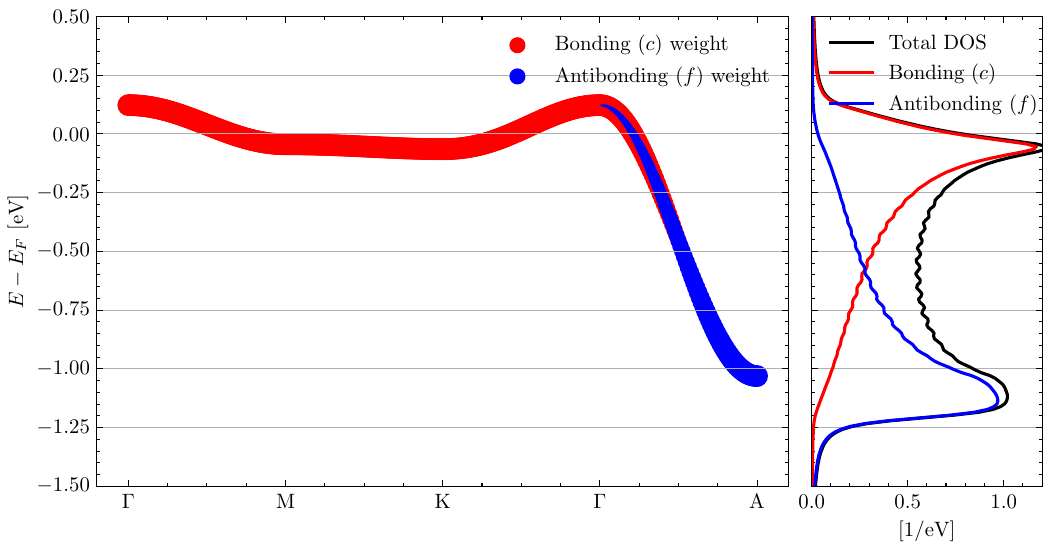}
        \caption{Left panel: Orbital character of the CMO band in the bonding-antibonding basis. Right panel: $c/f$ contribution to the local DOS.}
        \label{stacked_triangle_fat_band_dos}
\end{figure}

Figure~\ref{stacked_triangle_fat_band_dos} depicts the bonding-antibonding character of our effective single-orbital model on the stacked triangular lattice. The partially filled flat band portion and corresponding local DOS (see right panel) is dominated by the bonding ($c$) weight. Since the Fermi level of the original Ni$_3$In system resides near the peak of this partial density of states, our two-orbital decomposition yields an almost completely filled anti-bonding orbital, and a partially filled ($\sim 70$\%) bonding orbital. 

Since the correlated $c$-electrons are quite far from half-filling and the $f$ electrons are almost band insulating, the mapping in the Ni$_3$In case is not onto a standard ferromagnetic Kondo lattice (as in the case of other correlated electron compounds in the strange metal regime~\cite{hidden_kondo_werner}). The $c$ electrons are however clearly more correlated than the $f$  electrons and play the key role in the NFL behavior. 
In particular, sharp peaks in the DOS at the edge of the band, as in the case of the $c$-DOS, are known to favor ferromagnetic spin correlations~\cite{Ulmke1998}. This could be the reason why Ni$_3$In exhibits spin-freezing-like behavior and a linear-in-$T$ resistivity, reminiscent of Hund metals~\cite{georges_hunds_metals} and other strange metals in a spin-freezing crossover regime.  

\section{Conclusion}
\label{sec_conc}

We have studied the emergence of strange metallicity in the Kagome metal Ni$_3$In within DMFT applied to an effective one-band Hubbard model, where the quasi-flat band around the Fermi level is formed by compact molecular orbital states. This minimal model is sufficient to reproduce non-Fermi-liquid behavior, which manifests itself in the square-root like scaling of the imaginary part of the local self-energy, and in a weakly sub-linear $T$-dependence of the resistivity, as well as in the presence of local magnetic moments down to the lowest accessible temperatures. 

Hole-doping the model away from the sharp DOS peak associated with the almost flat band results in a crossover to a renormalized Fermi liquid at higher temperatures, in accordance with the hypothesized quantum-critical phase diagram in Ref.~\cite{ni3in_nature}. The quasi-flat band in the vicinity of the Fermi level plays a crucial role in the NFL behavior of Ni$_3$In. Bonding/anti-bonding transformations of a two-site model reveal that the relevant flat band portion is dominated by the bonding-electron weight, and that the bonding electrons exhibit a sharp peak at the upper edge of their partial DOS. Since the chemical potential is near this DOS peak, this leads to strong (likely ferromagnetic) spin correlations and spin-freezing related phenomena. Emerging NFL behavior in a single, partially-flat band Hubbard model with a qualitatively similar DOS structure has also been reported in Ref.~\cite{sayyad_nfl_one_band}. 

While ferromagnetism is a potential candidate, further investigations are required to uncover the nature of the hypothesized ordered state of electron-doped Ni$_3$In, as well as the role of other bands, such as those that form Dirac nodal rings in the vicinity of the flat band. The latter are believed to be responsible for the peak-dip structure of the differential conductance, as argued in Ref.~\cite{ni3in_origin_nature}. \newline

\acknowledgements

The calculations have been performed on the beo05 cluster at the University of Fribourg. We acknowledge support from the Swiss National Science Foundation via NCCR Marvel.

\appendix

\begin{widetext}
    
\section{Resistivity benchmark and uncertainty analysis}

\subsection{Resistivity benchmark for the Hubbard model at weak coupling}
\label{res_benchmark}

In this section, we benchmark our approach for the calculation of the DMFT resistivity, described in Section~\ref{sec_res_dmft}, on the square-lattice single-orbital Hubbard model. We consider quarter-filling and a small interaction $U = 2t$, where the system is in the Fermi liquid phase. In this regime, the resistivity is expected to scale as $\rho = A T^2$~\cite{fermi_hubbard_weak_coupling,2d_hubbard_jps}, where the coefficient $A$ is of the following form~\cite{transport_dmft_labollita}:
\begin{equation}
    A = \frac{24 C}{\Phi(\epsilon_F)} ,
    \label{a_coef}
\end{equation}
with $C$ being the quadratic coefficient of the imaginary part of the self-energy,
\begin{equation}
    |\text{Im} \Sigma(\omega)| = C(\omega^2 + \pi^2 T^2),
    \label{self_en_c}
\end{equation}
and $\Phi(\epsilon_F)$ the transport function evaluated at the Fermi energy. The transport function is defined as
\begin{equation}
    \Phi(\epsilon) = 2 \int_\text{BZ} \frac{d^2 k}{{(2\pi)}^2} (v_\kv^x)^2 \delta(\epsilon-\epsilon_\kv),
    \label{transport_func}
\end{equation}
with $\epsilon_\kv = -2t [ \cos(k_x) + \cos(k_y) ]$ the square lattice dispersion and $v_\kv$ the band velocity. Note the factor of 2, which accounts for spin degeneracy. 
Note that at half-filling, due to a perfectly square-like Fermi surface, the phase space for scattering is dominated by the nested states, resulting in a linear scaling of the resistivity with temperature~\cite{fermi_hubbard_weak_coupling,2d_hubbard_jps}. 

We now proceed to derive an analytic expression for $C$, in the small-coupling limit of the DMFT self-energy. For small $U \ll W$, where $W=8t$ is the bandwidth, the (spinless) self-energy can be approximated by a second order term~\cite{kotliar_ipt,fabrizio_many_body}:
\begin{equation}
    \Sigma(\omega) \simeq U n + \Sigma^{(2)}(\omega),
\end{equation}
where
\begin{equation}
    \begin{aligned}
        \Sigma^{(2)}(\omega) =& \, U^2 \int_{-\infty}^{0} d\epsilon_1 \int_{0}^{\infty} d\epsilon_2  \int_{0}^{\infty} d\epsilon_3 \frac{D(\epsilon_1)D(\epsilon_2) D(\epsilon_3)}{\omega + \epsilon_1 - \epsilon_2 - \epsilon_3 + i 0^+} 
        \\ &+ U^2 \int_{0}^{\infty} d\epsilon_1 \int_{-\infty}^{0} d\epsilon_2  \int_{-\infty}^{0} d\epsilon_3 \frac{D(\epsilon_1) D(\epsilon_2) D(\epsilon_3)}{\omega + \epsilon_1 - \epsilon_2 - \epsilon_3 + i 0^+}.
    \end{aligned}
\end{equation}
Here, $D(\epsilon)$ can be taken as the non-interacting density of states (per spin), with $\epsilon=0$ corresponding to the DOS at the chemical potential. We will focus on the second order term, which is frequency-dependent, and which eventually gives rise to the quadratic coefficient $C$ (the term which is linear in $U$ can be absorbed into a shift of the chemical potential). Making use of the Cauchy principal value identity, the imaginary part of the second order self-energy reads
\begin{equation}
    \begin{aligned}
        \text{Im} \Sigma(\omega) = &- \pi U^2 \int_{-\infty}^{0} d\epsilon_1 \int_{0}^{\infty} d\epsilon_2  \int_{0}^{\infty} d\epsilon_3 D(\epsilon_1)D(\epsilon_2) D(\epsilon_3) \delta(\omega + \epsilon_1 - \epsilon_2 - \epsilon_3) \\ &- \pi U^2 \int_{0}^{\infty} d\epsilon_1 \int_{-\infty}^{0} d\epsilon_2  \int_{-\infty}^{0} d\epsilon_3 D(\epsilon_1)D(\epsilon_2) D(\epsilon_3) \delta(\omega + \epsilon_1 - \epsilon_2 - \epsilon_3).
    \end{aligned}
\end{equation}
For $\omega \geq 0$, only the first line contributes, and hence the integral reduces to
\begin{equation}
    \begin{aligned}
        \text{Im} \Sigma(\omega) &= - \pi U^2 \int_{-\infty}^{0} d\epsilon_1 \int_{0}^{\infty} d\epsilon_2  \int_{0}^{\infty} d\epsilon_3 D(\epsilon_1)D(\epsilon_2) D(\epsilon_3) \delta(\omega + \epsilon_1 - \epsilon_2 - \epsilon_3) .
    \end{aligned}
\end{equation}
Integrating over $\epsilon_3$ yields
\begin{equation}
    \begin{aligned}
        \text{Im} \Sigma(\omega) &= - \pi U^2 \int_{-\infty}^{0} d\epsilon_1 \int_{0}^{\infty} d\epsilon_2 D(\epsilon_1)D(\epsilon_2) D(\omega + \epsilon_1 - \epsilon_2) \Theta(\omega + \epsilon_1 - \epsilon_2 \geq 0) ,
    \end{aligned}
\end{equation}
where $\Theta(x)$ is the Heaviside step-function. We note that our integral is confined to a triangle of height and width $\omega$ in the $\epsilon_1-\epsilon_2$ plane. For small frequencies $\omega$, we can approximate the densities of states by their values at the chemical potential:
\begin{equation}
    \begin{aligned}
        \text{Im} \Sigma(\omega) &\simeq - \pi U^2 D(0)^3 \int_{-\infty}^{0} d\epsilon_1 \int_{0}^{\infty} d\epsilon_2  \Theta(\omega + \epsilon_1 - \epsilon_2 \geq 0) \\ &= - \frac{\pi U^2 D{(0)}^3 \omega^2}{2},
    \end{aligned}
\end{equation}
where $\omega^2/2$ is the area of the effective triangle. Comparing the $\omega^2$ coefficient to the one in Eq.~\eqref{self_en_c}, yields
\begin{equation}
    C = \frac{\pi U^2 D{(0)}^3}{2} .
\end{equation}
The transport function in Eq.~\eqref{transport_func} is evaluated numerically. We use a $1500\times1500$ $k$-point grid, and a broadening parameter $\eta=0.01$ for the Gaussian representation of the Dirac delta functions. 

Figure~\ref{fig_benchmark} depicts the numerically computed resistivity vs the analytical form, with the coefficient $A$ from Eq.~\eqref{a_coef}. The good agreement between the two demonstrates the viability of the numerical approach outlined in Section~\ref{sec_res_dmft}.

\begin{figure}[H]
    \centering
    \includegraphics[width=0.4\textwidth]{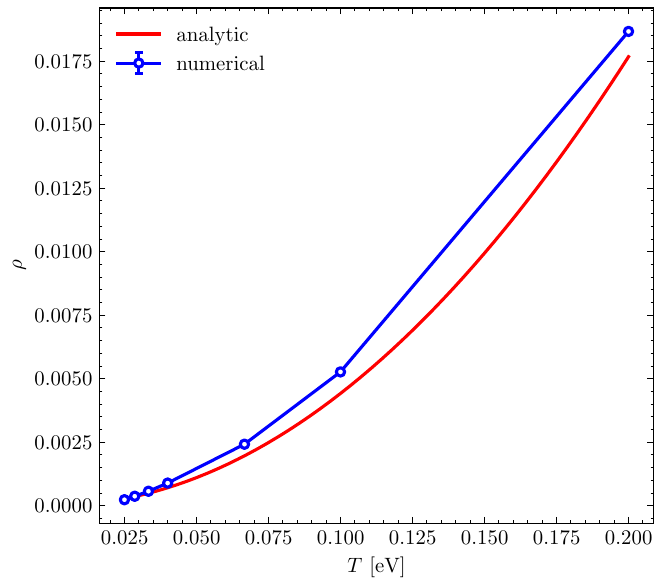}
    \caption{Resistivity benchmark for the square-lattice Hubbard model at quarter-filling and weak-coupling $U=2t$. The error bars are smaller than the plot markers.
    }
    \label{fig_benchmark}
\end{figure}

\subsection{Uncertainty of the fitting procedure}
\label{uncertainty_fit}

Here we explain how we quantified the fitting error in the DC resistivity, which is reflected in the vertical error bars of Fig.~\ref{rho_t_doped} and Fig.~\ref{fig_benchmark}. For this, we recall that the DC resistivity is obtained from the ratio of the fitted parameters: $\rho_\text{DC} = \frac{\Gamma}{D}$. The uncertainty in the $(D,\Gamma)$ parameters is given by the covariance matrix of the \texttt{curve\_fit} routine in the Python SciPy library~\cite{scipy}. The variance of the static resistivity, which is a function of $(D,\Gamma)$ is given by~\cite{statproofbook}
\begin{equation}
    \begin{aligned}
        \text{Var}[\rho_\text{DC}] &= \Big( \frac{\partial \rho_\text{DC}}{\partial D}\Big)^2 \text{Var}[D] + \Big( \frac{\partial \rho_\text{DC}}{\partial \Gamma}\Big)^2 \text{Var}[\Gamma] + 2 \frac{\partial\rho_\text{DC} }{\partial D}\frac{\partial\rho_\text{DC} }{\partial \Gamma} \text{Cov}[D,\Gamma] \\ &=  \frac{\Gamma^2}{D^4} \text{Var}[D] + \frac{1}{D^2} \text{Var}[\Gamma] - 2 \frac{\Gamma}{D^3}\text{Cov}[D,\Gamma],
    \end{aligned}
\end{equation}
where the variance (covariance) of $D$ and $\Gamma$ is given by the diagonal (off-diagonal) element of the covariance matrix estimated by the \texttt{curve\_fit} routine. The absolute error is then given by $\Delta \rho_\text{DC} = \sqrt{\text{Var}[\rho_\text{DC}]}$.

\end{widetext}

\bibliography{ni3in}

@article{Werner2006,
  title = {Continuous-Time Solver for Quantum Impurity Models},
  author = {Werner, Philipp and Comanac, Armin and de' Medici, Luca and Troyer, Matthias and Millis, Andrew J.},
  journal = {Phys. Rev. Lett.},
  volume = {97},
  issue = {7},
  pages = {076405},
  numpages = {4},
  year = {2006},
  month = {Aug},
  publisher = {American Physical Society},
  doi = {10.1103/PhysRevLett.97.076405},
  url = {https://link.aps.org/doi/10.1103/PhysRevLett.97.076405}
}

@article{Sayyad2020,
  title = {Pairing and non-Fermi liquid behavior in partially flat-band systems: Beyond nesting physics},
  author = {Sayyad, Sharareh and Huang, Edwin W. and Kitatani, Motoharu and Vaezi, Mohammad-Sadegh and Nussinov, Zohar and Vaezi, Abolhassan and Aoki, Hideo},
  journal = {Phys. Rev. B},
  volume = {101},
  issue = {1},
  pages = {014501},
  numpages = {10},
  year = {2020},
  month = {Jan},
  publisher = {American Physical Society},
  doi = {10.1103/PhysRevB.101.014501},
  url = {https://link.aps.org/doi/10.1103/PhysRevB.101.014501}
}

@article{Werner2016,
  title = {Spin-freezing perspective on cuprates},
  author = {Werner, Philipp and Hoshino, Shintaro and Shinaoka, Hiroshi},
  journal = {Phys. Rev. B},
  volume = {94},
  issue = {24},
  pages = {245134},
  numpages = {9},
  year = {2016},
  month = {Dec},
  publisher = {American Physical Society},
  doi = {10.1103/PhysRevB.94.245134},
  url = {https://link.aps.org/doi/10.1103/PhysRevB.94.245134}
}

@article{av3sb5_kagome,
  title = {New kagome prototype materials: discovery of ${\mathrm{KV}}_{3}{\mathrm{Sb}}_{5},{\mathrm{RbV}}_{3}{\mathrm{Sb}}_{5}$, and ${\mathrm{CsV}}_{3}{\mathrm{Sb}}_{5}$},
  author = {Ortiz, Brenden R. and Gomes, L\'{\i}dia C. and Morey, Jennifer R. and Winiarski, Michal and Bordelon, Mitchell and Mangum, John S. and Oswald, Iain W. H. and Rodriguez-Rivera, Jose A. and Neilson, James R. and Wilson, Stephen D. and Ertekin, Elif and McQueen, Tyrel M. and Toberer, Eric S.},
  journal = {Phys. Rev. Mater.},
  volume = {3},
  issue = {9},
  pages = {094407},
  numpages = {9},
  year = {2019},
  month = {Sep},
  publisher = {American Physical Society},
  doi = {10.1103/PhysRevMaterials.3.094407},
  url = {https://link.aps.org/doi/10.1103/PhysRevMaterials.3.094407}
}

@article{kv3sb5_cdw,
  author = {Luo, Hailan and Gao, Qiang and Liu, Hongxiong and Gu, Yuhao and Wu, Dingsong and Yi, Changjiang and Jia, Junjie and Wu, Shilong and Luo, Xiangyu and Xu, Yu and Zhao, Lin and Wang, Qingyan and Mao, Hanqing and Liu, Guodong and Zhu, Zhihai and Shi, Youguo and Jiang, Kun and Hu, Jiangping and Xu, Zuyan and Zhou, X. J.},
  title = {Electronic nature of charge density wave and electron-phonon coupling in kagome superconductor KV3Sb5},
  journal = {Nature Communications},
  year = {2022},
  volume = {13},
  number = {1},
  pages = {273},
  doi = {10.1038/s41467-021-27946-6},
  url = {https://doi.org/10.1038/s41467-021-27946-6},
  issn = {2041-1723}
}

@article{fege_cdw,
  author = {Teng, Xiaokun and Chen, Lebing and Ye, Feng and Rosenberg, Elliott and Liu, Zhaoyu and Yin, Jia-Xin and Jiang, Yu-Xiao and Oh, Ji Seop and Hasan, M. Zahid and Neubauer, Kelly J. and Gao, Bin and Xie, Yaofeng and Hashimoto, Makoto and Lu, Donghui and Jozwiak, Chris and Bostwick, Aaron and Rotenberg, Eli and Birgeneau, Robert J. and Chu, Jiun-Haw and Yi, Ming and Dai, Pengcheng},
  title = {Discovery of charge density wave in a kagome lattice antiferromagnet},
  journal = {Nature},
  year = {2022},
  volume = {609},
  number = {7927},
  pages = {490--495},
  doi = {10.1038/s41586-022-05034-z},
  url = {https://doi.org/10.1038/s41586-022-05034-z},
  issn = {1476-4687}
}

@article{fege_cdw_mag,
  author = {Teng, Xiaokun and Oh, Ji Seop and Tan, Hengxin and Chen, Lebing and Huang, Jianwei and Gao, Bin and Yin, Jia-Xin and Chu, Jiun-Haw and Hashimoto, Makoto and Lu, Donghui and Jozwiak, Chris and Bostwick, Aaron and Rotenberg, Eli and Granroth, Garrett E. and Yan, Binghai and Birgeneau, Robert J. and Dai, Pengcheng and Yi, Ming},
  title = {Magnetism and charge density wave order in kagome FeGe},
  journal = {Nature Physics},
  year = {2023},
  volume = {19},
  number = {6},
  pages = {814--822},
  doi = {10.1038/s41567-023-01985-w},
  url = {https://doi.org/10.1038/s41567-023-01985-w},
  issn = {1745-2481}
}

@article{kagome_magnets_nature,
  author = {Yin, Jia-Xin and Lian, Biao and Hasan, M. Zahid},
  title = {Topological kagome magnets and superconductors},
  journal = {Nature},
  year = {2022},
  volume = {612},
  number = {7941},
  pages = {647--657},
  doi = {10.1038/s41586-022-05516-0},
  url = {https://doi.org/10.1038/s41586-022-05516-0},
  issn = {1476-4687}
}

@article{thomale_kagome,
  title = {Unconventional Fermi Surface Instabilities in the Kagome Hubbard Model},
  author = {Kiesel, Maximilian L. and Platt, Christian and Thomale, Ronny},
  journal = {Phys. Rev. Lett.},
  volume = {110},
  issue = {12},
  pages = {126405},
  numpages = {5},
  year = {2013},
  month = {Mar},
  publisher = {American Physical Society},
  doi = {10.1103/PhysRevLett.110.126405},
  url = {https://link.aps.org/doi/10.1103/PhysRevLett.110.126405}
}

@article{topology_sc_kagome,
  author = {Yin, Jia-Xin and Lian, Biao and Hasan, M. Zahid},
  title = {Topological kagome magnets and superconductors},
  journal = {Nature},
  year = {2022},
  volume = {612},
  number = {7941},
  pages = {647--657},
  doi = {10.1038/s41586-022-05516-0},
  url = {https://doi.org/10.1038/s41586-022-05516-0},
  issn = {1476-4687}
}

@article{sc_cr_based_kagome,
  author = {Liu, Yi and Liu, Zi-Yi and Bao, Jin-Ke and Yang, Peng-Tao and Ji, Liang-Wen and Wu, Si-Qi and Shen, Qin-Xin and Luo, Jun and Yang, Jie and Liu, Ji-Yong and Xu, Chen-Chao and Yang, Wu-Zhang and Chai, Wan-Li and Lu, Jia-Yi and Liu, Chang-Chao and Wang, Bo-Sen and Jiang, Hao and Tao, Qian and Ren, Zhi and Xu, Xiao-Feng and Cao, Chao and Xu, Zhu-An and Zhou, Rui and Cheng, Jin-Guang and Cao, Guang-Han},
  title = {Superconductivity under pressure in a chromium-based kagome metal},
  journal = {Nature},
  year = {2024},
  volume = {632},
  number = {8027},
  pages = {1032--1037},
  doi = {10.1038/s41586-024-07761-x},
  url = {https://doi.org/10.1038/s41586-024-07761-x},
  issn = {1476-4687}
}

@article{ni3in_nature,
author={Ye, Linda
and Fang, Shiang
and Kang, Mingu
and Kaufmann, Josef
and Lee, Yonghun
and John, Caolan
and Neves, Paul M.
and Zhao, S. Y. Frank
and Denlinger, Jonathan
and Jozwiak, Chris
and Bostwick, Aaron
and Rotenberg, Eli
and Kaxiras, Efthimios
and Bell, David C.
and Janson, Oleg
and Comin, Riccardo
and Checkelsky, Joseph G.},
title={Hopping frustration-induced flat band and strange metallicity in a kagome metal},
journal={Nature Physics},
year={2024},
month={Apr},
day={01},
volume={20},
number={4},
pages={610-614},
issn={1745-2481},
doi={10.1038/s41567-023-02360-5},
url={https://doi.org/10.1038/s41567-023-02360-5}
}

@article{ni3in_origin_nature,
  author = {Souza, Jean C. and Haim, Moshe and Gupta, Ambikesh and Mahankali, Mounica and Xie, Fang and Fang, Yuan and Chen, Lei and Fang, Shiang and Tan, Hengxin and Han, Minyong and John, Caolan and Zheng, Jingxu and Liu, Yiwen and Yan, Binghai and Checkelsky, Joseph G. and Si, Qimiao and Avraham, Nurit and Beidenkopf, Haim},
  title = {Origin of strange metallicity in a d-orbital kagome metal},
  journal = {Nature Physics},
  year = {2026},
  doi = {10.1038/s41567-026-03216-4},
  url = {https://doi.org/10.1038/s41567-026-03216-4},
  issn = {1745-2481}
}

@misc{ni3in_cmo,
      title={Correlated flat-band physics in a bilayer kagome metal based on compact molecular orbitals}, 
      author={Mounica Mahankali and Fang Xie and Yuan Fang and Lei Chen and Shouvik Sur and Silke Paschen and Jean C. Souza and Moshe Haim and Ambikesh Gupta and Nurit Avraham and Haim Beidenkopf and Hengxin Tan and Binghai Yan and Qimiao Si},
      year={2025},
      eprint={2503.09706},
      archivePrefix={arXiv},
      primaryClass={cond-mat.str-el},
      url={https://arxiv.org/abs/2503.09706}, 
}

@article{strange_metals_perspective_nature,
  author = {Checkelsky, Joseph G. and Bernevig, B. Andrei and Coleman, Piers and Si, Qimiao and Paschen, Silke},
  title = {Flat bands, strange metals and the Kondo effect},
  journal = {Nature Reviews Materials},
  year = {2024},
  volume = {9},
  number = {7},
  pages = {509--526},
  doi = {10.1038/s41578-023-00644-z},
  url = {https://doi.org/10.1038/s41578-023-00644-z},
  issn = {2058-8437}
}

@article{resistivity_la2cuo4_science,
author = {R. A. Cooper  and Y. Wang  and B. Vignolle  and O. J. Lipscombe  and S. M. Hayden  and Y. Tanabe  and T. Adachi  and Y. Koike  and M. Nohara  and H. Takagi  and Cyril Proust  and N. E. Hussey },
title = {Anomalous Criticality in the Electrical Resistivity of $\mathrm{La}_{2-x}\mathrm{Sr}_x\mathrm{CuO}_4$},
journal = {Science},
volume = {323},
number = {5914},
pages = {603-607},
year = {2009},
doi = {10.1126/science.1165015},
URL = {https://www.science.org/doi/10.1126/science.1165015},
eprint = {https://www.science.org/doi/10.1126/science.1165015}
}

@article{resistivity_heavy_fermion_nature,
  author = {Nguyen, D. H. and Sidorenko, A. and Taupin, M. and Knebel, G. and Lapertot, G. and Schuberth, E. and Paschen, S.},
  title = {Superconductivity in an extreme strange metal},
  journal = {Nature Communications},
  year = {2021},
  volume = {12},
  number = {1},
  pages = {4341},
  doi = {10.1038/s41467-021-24670-z},
  url = {https://doi.org/10.1038/s41467-021-24670-z},
  issn = {2041-1723}
}

@article{matbg_strange_metallicity,
  author = {Jaoui, Alexandre and Das, Ipsita and Di Battista, Giorgio and Díez-Mérida, Jaime and Lu, Xiaobo and Watanabe, Kenji and Taniguchi, Takashi and Ishizuka, Hiroaki and Levitov, Leonid and Efetov, Dmitri K.},
  title = {Quantum critical behaviour in magic-angle twisted bilayer graphene},
  journal = {Nature Physics},
  year = {2022},
  volume = {18},
  number = {6},
  pages = {633--638},
  doi = {10.1038/s41567-022-01556-5},
  url = {https://doi.org/10.1038/s41567-022-01556-5},
  issn = {1745-2481}
}

@article{ni3in_prx,
  title = {Topological Flat-Band-Driven Metallic Thermoelectricity},
  author = {Garmroudi, Fabian and Coulter, Jennifer and Serhiienko, Illia and Di Cataldo, Simone and Parzer, Michael and Riss, Alexander and Grasser, Matthias and Stockinger, Simon and Khmelevskyi, Sergii and Pryga, Kacper and Wiendlocha, Bartlomiej and Held, Karsten and Mori, Takao and Bauer, Ernst and Georges, Antoine and Pustogow, Andrej},
  journal = {Phys. Rev. X},
  volume = {15},
  issue = {2},
  pages = {021054},
  numpages = {11},
  year = {2025},
  month = {May},
  publisher = {American Physical Society},
  doi = {10.1103/PhysRevX.15.021054},
  url = {https://link.aps.org/doi/10.1103/PhysRevX.15.021054}
}

@article{dmft_georges,
  title = {Dynamical mean-field theory of strongly correlated fermion systems and the limit of infinite dimensions},
  author = {Georges, Antoine and Kotliar, Gabriel and Krauth, Werner and Rozenberg, Marcelo J.},
  journal = {Rev. Mod. Phys.},
  volume = {68},
  issue = {1},
  pages = {13--125},
  numpages = {0},
  year = {1996},
  month = {Jan},
  publisher = {American Physical Society},
  doi = {10.1103/RevModPhys.68.13},
  url = {https://link.aps.org/doi/10.1103/RevModPhys.68.13}
}

@article{conductivity_hubbard_prl,
  title = {Conductivity in the Square Lattice Hubbard Model at High Temperatures: Importance of Vertex Corrections},
  author = {Vu\ifmmode \check{c}\else \v{c}\fi{}i\ifmmode \check{c}\else \v{c}\fi{}evi\ifmmode \acute{c}\else \'{c}\fi{}, J. and Kokalj, J. and \ifmmode \check{Z}\else \v{Z}\fi{}itko, R. and Wentzell, N. and Tanaskovi\ifmmode \acute{c}\else \'{c}\fi{}, D. and Mravlje, J.},
  journal = {Phys. Rev. Lett.},
  volume = {123},
  issue = {3},
  pages = {036601},
  numpages = {6},
  year = {2019},
  month = {Jul},
  publisher = {American Physical Society},
  doi = {10.1103/PhysRevLett.123.036601},
  url = {https://link.aps.org/doi/10.1103/PhysRevLett.123.036601}
}

@article{local_spin_dyn_mit,
  title = {Quantum critical local spin dynamics near the Mott metal-insulator transition in infinite dimensions},
  author = {Dasari, Nagamalleswararao and Vidhyadhiraja, N. S. and Jarrell, Mark and McKenzie, Ross H.},
  journal = {Phys. Rev. B},
  volume = {95},
  issue = {16},
  pages = {165105},
  numpages = {8},
  year = {2017},
  month = {Apr},
  publisher = {American Physical Society},
  doi = {10.1103/PhysRevB.95.165105},
  url = {https://link.aps.org/doi/10.1103/PhysRevB.95.165105}
}

@article{sampling_ct_hyb_kowalski,
  title = {State and superstate sampling in hybridization-expansion continuous-time quantum Monte Carlo},
  author = {Kowalski, Alexander and Hausoel, Andreas and Wallerberger, Markus and Gunacker, Patrik and Sangiovanni, Giorgio},
  journal = {Phys. Rev. B},
  volume = {99},
  issue = {15},
  pages = {155112},
  numpages = {14},
  year = {2019},
  month = {Apr},
  publisher = {American Physical Society},
  doi = {10.1103/PhysRevB.99.155112},
  url = {https://link.aps.org/doi/10.1103/PhysRevB.99.155112}
}

@article{nfl_nature_werner,
  author    = {Philipp Werner and Michele Casula and Takashi Miyake and Ferdi Aryasetiawan and Andrew J. Millis and Silke Biermann},
  title     = {Satellites and large doping and temperature dependence of electronic properties in hole-doped BaFe$_2$As$_2$},
  journal   = {Nature Physics},
  year      = {2012},
  volume    = {8},
  number    = {4},
  pages     = {331--337},
  month     = {4},
  doi       = {10.1038/nphys2250},
  url       = {https://doi.org/10.1038/nphys2250},
  issn      = {1745-2481}
}

@article{hidden_kondo_werner,
  title = {Hidden Kondo lattice physics in single-orbital Hubbard models},
  author = {Werner, Philipp and Ghorashi, Sayed Ali Akbar},
  journal = {Phys. Rev. B},
  volume = {111},
  issue = {4},
  pages = {045138},
  numpages = {11},
  year = {2025},
  month = {Jan},
  publisher = {American Physical Society},
  doi = {10.1103/PhysRevB.111.045138},
  url = {https://link.aps.org/doi/10.1103/PhysRevB.111.045138}
}

@article{sayyad_nfl_one_band,
  title = {Pairing and non-Fermi liquid behavior in partially flat-band systems: Beyond nesting physics},
  author = {Sayyad, Sharareh and Huang, Edwin W. and Kitatani, Motoharu and Vaezi, Mohammad-Sadegh and Nussinov, Zohar and Vaezi, Abolhassan and Aoki, Hideo},
  journal = {Phys. Rev. B},
  volume = {101},
  issue = {1},
  pages = {014501},
  numpages = {10},
  year = {2020},
  month = {Jan},
  publisher = {American Physical Society},
  doi = {10.1103/PhysRevB.101.014501},
  url = {https://link.aps.org/doi/10.1103/PhysRevB.101.014501}
}

@article{prl_nfl_shinaoka,
  title = {Phase Diagram of Pyrochlore Iridates: All-in--All-out Magnetic Ordering and Non-Fermi-Liquid Properties},
  author = {Shinaoka, Hiroshi and Hoshino, Shintaro and Troyer, Matthias and Werner, Philipp},
  journal = {Phys. Rev. Lett.},
  volume = {115},
  issue = {15},
  pages = {156401},
  numpages = {5},
  year = {2015},
  month = {Oct},
  publisher = {American Physical Society},
  doi = {10.1103/PhysRevLett.115.156401},
  url = {https://link.aps.org/doi/10.1103/PhysRevLett.115.156401}
}

@article{Ulmke1998,
  author    = {Ulmke, M.},
  title     = {Ferromagnetism in the Hubbard model on fcc-type lattices},
  journal   = {The European Physical Journal B - Condensed Matter and Complex Systems},
  year      = {1998},
  volume    = {1},
  number    = {3},
  pages     = {301--304},
  doi       = {10.1007/s100510050186},
  url       = {https://doi.org/10.1007/s100510050186},
  issn      = {1434-6036}
}

@article{georges_hunds_metals,
   author = "Georges, Antoine and Medici, Luca de&apos; and Mravlje, Jernej",
   title = "Strong Correlations from Hund’s Coupling", 
   journal= "Annual Review of Condensed Matter Physics",
   year = "2013",
   volume = "4",
   number = "Volume 4, 2013",
   pages = "137-178",
   doi = "https://doi.org/10.1146/annurev-conmatphys-020911-125045",
   url = "https://www.annualreviews.org/content/journals/10.1146/annurev-conmatphys-020911-125045",
   publisher = "Annual Reviews",
   issn = "1947-5462",
   type = "Journal Article",
  }

@article{fermi_hubbard_weak_coupling,
  title = {Transport in the two-dimensional Fermi-Hubbard model: Lessons from weak coupling},
  author = {Kiely, Thomas G. and Mueller, Erich J.},
  journal = {Phys. Rev. B},
  volume = {104},
  issue = {16},
  pages = {165143},
  numpages = {12},
  year = {2021},
  month = {Oct},
  publisher = {American Physical Society},
  doi = {10.1103/PhysRevB.104.165143},
  url = {https://link.aps.org/doi/10.1103/PhysRevB.104.165143}
}

@article{2d_hubbard_jps,
  title={Temperature dependence of electrical resistivity in two-dimensional fermi systems},
  author={Fujimoto, Satoshi and Kohno, Hiroshi and Yamada, Kosaku},
  journal={Journal of the Physical Society of Japan},
  volume={60},
  number={8},
  pages={2724--2728},
  year={1991},
  publisher={The Physical Society of Japan}
}

@article{transport_dmft_labollita,
   title={Low-temperature transport in high-conductivity correlated metals: A density functional plus dynamical mean-field study of cubic perovskites},
   volume={113},
   ISSN={2469-9969},
   url={http://dx.doi.org/10.1103/71c6-sb7v},
   DOI={10.1103/71c6-sb7v},
   number={8},
   journal={Physical Review B},
   publisher={American Physical Society (APS)},
   author={LaBollita, Harrison and Lee-Hand, Jeremy and Kugler, Fabian B. and Van Muñoz, Lorenzo and Beck, Sophie and Hampel, Alexander and Kaye, Jason and Georges, Antoine and Dreyer, Cyrus E.},
   year={2026},
   month=feb 
}

@article{kotliar_ipt,
  title = {New Iterative Perturbation Scheme for Lattice Models with Arbitrary Filling},
  author = {Kajueter, Henrik and Kotliar, Gabriel},
  journal = {Phys. Rev. Lett.},
  volume = {77},
  issue = {1},
  pages = {131--134},
  numpages = {0},
  year = {1996},
  month = {Jul},
  publisher = {American Physical Society},
  doi = {10.1103/PhysRevLett.77.131},
  url = {https://link.aps.org/doi/10.1103/PhysRevLett.77.131}
}

@book{fabrizio_many_body,
  author    = {Michele Fabrizio},
  title     = {A Course in Quantum Many-Body Theory: From Conventional Fermi Liquids to Strongly Correlated Systems},
  publisher = {Springer International Publishing},
  address   = {Cham},
  year      = {2022},
  isbn      = {978-3-031-16305-0},
  doi       = {10.1007/978-3-031-16305-0},
  url       = {https://doi.org/10.1007/978-3-031-16305-0}
}

@article{scipy,
  author  = {Virtanen, Pauli and Gommers, Ralf and Oliphant, Travis E. and
            Haberland, Matt and Reddy, Tyler and Cournapeau, David and
            Burovski, Evgeni and Peterson, Pearu and Weckesser, Warren and
            Bright, Jonathan and {van der Walt}, St{\'e}fan J. and
            Brett, Matthew and Wilson, Joshua and Millman, K. Jarrod and
            Mayorov, Nikolay and Nelson, Andrew R. J. and Jones, Eric and
            Kern, Robert and Larson, Eric and Carey, C J and
            Polat, {\.I}lhan and Feng, Yu and Moore, Eric W. and
            {VanderPlas}, Jake and Laxalde, Denis and Perktold, Josef and
            Cimrman, Robert and Henriksen, Ian and Quintero, E. A. and
            Harris, Charles R. and Archibald, Anne M. and
            Ribeiro, Ant{\^o}nio H. and Pedregosa, Fabian and
            {van Mulbregt}, Paul and {SciPy 1.0 Contributors}},
  title   = {{{SciPy} 1.0: Fundamental Algorithms for Scientific
            Computing in Python}},
  journal = {Nature Methods},
  year    = {2020},
  volume  = {17},
  pages   = {261--272},
  adsurl  = {https://rdcu.be/b08Wh},
  doi     = {10.1038/s41592-019-0686-2},
}

@misc{statproofbook,
  author       = {Joram Soch and Karahan Sarıtaş and Maja and Pietro Monticone and Thomas J. Faulkenberry and Eric Pedersen and Heiner Atze and Osvaldo A Martin and Alex Kipnis and Salvador Balkus and lfkdlfdlk and Alexander D. Bolton and Adam Knapp and Carsten Allefeld and Ciarán D. McInerney and Lo4ding00 and Luiz Max Carvalho and Mario5572 and maxgrozo},
  title        = {StatProofBook/StatProofBook.github.io: StatProofBook 2025 (Version 2025)},
  year         = {2025},
  publisher    = {Zenodo},
  doi          = {10.5281/zenodo.18096132},
  url          = {https://doi.org/10.5281/zenodo.18096132}
}

\end{document}